\documentclass[aps,11pt,preprintnumbers,nofootinbib]{revtex4}
\usepackage{graphicx}
\usepackage{amssymb,amsmath}
\usepackage{slashed}
\usepackage{epsfig}
\usepackage{amsmath}
\usepackage{bm}
\usepackage{times}
\usepackage{color}
\def\nn{\nonumber}
\def\bea{\begin{eqnarray}}
\def\eea{\end{eqnarray}}
\def\ba{\begin{eqnarray}}
\def\ea{\end{eqnarray}}
\def\be{\begin{equation}}
\def\ee{\end{equation}}

\begin{document}
\preprint{Published in JHEP}
\title{\Large High $p_T$ Production of $b \, \bar{b}$ at LHC and New $\rm SU(3)_c$ Bosons}
\author{Bartosz Fornal$^{1,}$}
\email[]{ fornal@theory.caltech.edu}
\author{Michael Trott$^{2,}$}
\email[]{ mtrott@perimeterinstitute.ca}

\affiliation{$^{1}$ California Institute of Technology, Pasadena, CA 91125, USA}

\affiliation{$^{2}$ Perimeter Institute for Theoretical Physics, Waterloo, ON N2J-2W9, Canada.}

\begin{abstract}
We discuss the potential of measurements of $\sigma(p \, p \rightarrow b \, \bar{b})$
to constrain new bosonic degrees of freedom at the LHC when $p_T \gtrsim \frac{\sqrt{s}}{11} {\rm \, TeV}$ for a pseudorapidity cut $|\eta| < 2.4$. By suppressing the
NLO QCD production of $b \, \bar{b}$ pairs through simple kinematic constraints we show how to more
efficiently exploit CMS's reach out to $1.5 \, {\rm TeV}$ in $p_T$ for $B$ mesons in searches for new physics.
Using this technique we investigate the potential for discovery of new massive spin one and spin zero $\rm SU(3)_c$ octet bosons by analyzing their effect on the $d \sigma(p \, p \rightarrow b \, \bar{b})/d p_T^2$ distribution.
In some cases, the tree level contributions are small and the largest effect of new physics on the $p_T$ tail of the distribution is due to one-loop resonant $s$-channel production or interference effects with
the LO standard model production mechanisms $q \, \bar{q},  g \, g \rightarrow b \, \bar{b}$.
We explore this possibility in some detail when the standard model is extended with an $(8,2)_{1/2}$ scalar motivated by
Minimal Flavor Violation.
\end{abstract}
\maketitle
\section{Introduction}

The Large Hadron Collider (LHC) will measure the $b$ jet production cross section in hadron collisions for various ranges of transverse momenta ($p_T$) with unprecedented statistics and kinematics.
With the improvements of $b$ tagging pioneered at the Tevatron \cite{Kreps:2009nd,Moulik:2007px},
the analysis of $b$ production with associated muon tags in the LHC environment \cite{CMSnote,CMS2} is very promising.
Studies by the CMS collaboration have determined that the final states of $b$ production
can be triggered on and measured out to $1.5 \, {\rm TeV}$ in the $p_T$ of inclusive $B$ meson production by utilizing $b$ jet tagging in events containing at least one muon \cite{CMSnote}.
The standard model (SM) QCD production mechanism for $b \, \bar{b}$ falls rapidly
when $p_T > m_b$, thus, studies with restricted $p_T \gtrsim \frac{\sqrt{s}}{11} {\rm \, TeV}$ (for such values of $p_T$ and a pseudorapidity ($\eta$) of $|\eta| < 2.4$ the variables $p_T$ and $\eta$ are independent from one another)
offer the opportunity to explore the effect that beyond the SM (BSM) physics can have on the $p_T$ tail of the $d\sigma(p \, p \rightarrow b \, \bar{b})/d p_T^2$ distribution.

In utilizing this distribution to search for BSM physics one has to contend with the fact that
NLO QCD production leads to a larger background than LO QCD at high $p_T$ due to a $t$-channel singularity in $g\,g \rightarrow g\,g$. In this paper, we argue that the NLO QCD production background can be significantly reduced through simple
kinematic cuts. When the cuts are imposed, the dominant remaining QCD production processes are $q \, \bar{q},  g \, g \rightarrow b \, \bar{b}$
and BSM physics can more easily produce statistically significant excesses of high $p_T$
events.

We restrict our attention to examples of BSM physics with massive new octets under $\rm SU(3)_c$ and find that measurements of $d\sigma(p \, p \rightarrow b \, \bar{b})/d p_T^2$
are quite promising to constrain some $\rm SU(3)_c$ octets, affording a $5 \, \sigma$ discovery reach out to $m \lesssim 2 \, p_T^{\rm max}$ using these cuts.
The appearance of new bosons of the form discussed in this paper can be motivated in many scenarios of BSM physics.
We concentrate on two models
where they appear: a model when a larger gauge group
breaks down to QCD as a diagonal subgroup \cite{Preskill:1980mz,Hall:1985wz,Frampton:1987dn,Hill:1991at,Hill:1993hs} and the Manohar-Wise model \cite{Manohar} which considers
Minimal Flavor Violation (MFV) \cite{Chivukula:1987py,Hall:1990ac,D'Ambrosio:2002ex,Buras:2003jf} as a global symmetry of new physics that allows specific scalar representations \cite{Manohar, Arnold:2009ay}.

We first discuss isolating the signal region by suppressing the NLO QCD production mechanisms through simple kinematic cuts
and determine analytically the resulting LO QCD contribution to $d\sigma(p \, p \rightarrow b \, \bar{b})/d p_T^2$. We then
examine the effect on  the calculated distribution due to spin one $\rm SU(3)_c$ octets and determine the $5 \, \sigma$ discovery reach.\footnote{
In this paper we focus on the partonic production mechanisms of QCD and BSM physics. The produced
$b$ quarks in both cases can be related through common perturbative fragmentation functions to $B$ mesons
or other hadronic final states (or jet functions) and the relative event rates and discovery reach
conclusions will not be significantly effected by this further complication.}

The tree level effects of some new  $\rm SU(3)_c$ bosons can be small. This is the case if the strength of the boson's coupling to quarks is proportional to the quark mass.
In some models, the largest effect of new physics comes through loop corrections interfering with the SM production mechanisms  $q \, \bar{q},  g \, g \rightarrow b \, \bar{b}$ or
one-loop production that is enhanced with an $s$-channel resonance.
We explore this possibility in detail when the SM is supplemented with an $(8,2)_{1/2}$ scalar motivated by MFV \cite{Manohar}. However, for this model,
the potential for discovery in this channel is not strong.

\section{QCD production in the high $p_T$ region}\label{justifyQCD}
The dominant contribution to the $d\sigma(p \, p \rightarrow b \, \bar{b})/d p_T^2$ distribution is QCD production.
We first discuss our estimates of the QCD production mechanisms and the isolation of this signal before comparing the
BSM signal to this background.
When kinematic cuts that we advocate are imposed, the dominant
SM production mechanisms are $q \, \bar{q},  g \, g \rightarrow b \, \bar{b}$ and the BSM signature must be detectable
above this background.
The hadronic momenta are defined as $P_{1,2}^h = \frac{\sqrt{s}}{2} \, (1, \vec{0},\pm1)$, where $\sqrt{s}$ is the hadronic center of mass energy.
The momentum of the $b$ quark is defined in terms of $\eta,p_T$ by
\bea
p = \left(|p_T| \cosh\eta, \vec{p}_T, |p_T| \sinh\eta \right).
\eea

\begin{figure}[h] \label{LOQCD}
\begin{center}
{\includegraphics[height=7.9cm]{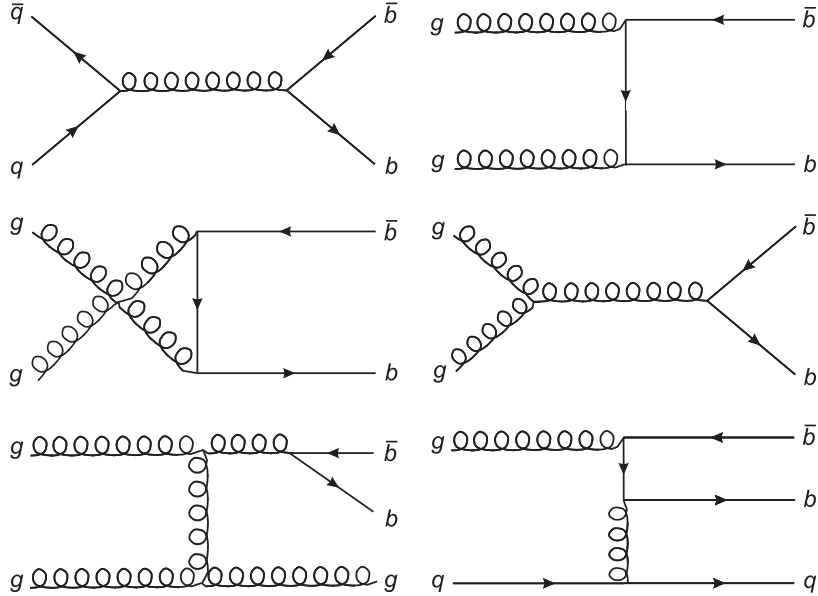}}
\caption{The LO heavy flavor production in the flavor creation (top two rows), gluon splitting (bottom left) and
flavor excitation (bottom right) channels.}
\end{center}
\end{figure}

For QCD, the production processes $q \, \bar{q},  g \, g \rightarrow b \, \bar{b}$ begin at $\mathcal{O}(\alpha_s^2)$ (see Figure 1).
Alternate heavy flavor production mechanisms through gluon splitting $g \, g \rightarrow gg \rightarrow b \, \bar{b} \, g$ and
direct heavy flavor excitation $g \, q \rightarrow b \, \bar{b} \, q$ begin at $\mathcal{O}(\alpha_s^3)$. The contributions from gluon splitting and flavor excitation are dominant when all phase space is integrated over.  This has been appreciated theoretically for some time, see \cite{Kunszt:1979ci,Ellis:1986ef}.
Studies of the total cross section at LHC (operating at $\sqrt{s} = 10 \, {\rm TeV}$)
using Pythia \cite{Nason:1999ta} find that  $\sigma(p \, p \rightarrow b \, \bar{b})$ is approximately $\sim 460 \, {\rm \mu b}$, with gluon splitting and heavy flavor excitation processes contributing
$\sim 200 \,  {\rm \mu b}$ each while the processes $q \, \bar{q},  g \, g \rightarrow b \, \bar{b}$  contribute
 $\sim 50  \, {\rm \mu b}$. This is mostly due to the $t$-channel Rutherford scattering enhancement at large $p_T$ for $g \, g \rightarrow g \, g$ and the effect of PDFs;  $\sigma(g\,g \rightarrow g \,g)$ is approximately a factor of 200 larger than
$\sigma(q \, \bar{q} \rightarrow b \, \bar{b})$ in the high $p_T$ region.

The different processes can be distinguished through their final state kinematics. The processes  $q \, \bar{q},  g \, g \rightarrow b \, \bar{b}$ yield $b \, \bar{b}$ pairs that tend to be
back to back and highly symmetric in $p_T$.  In flavor excitation, only one of the quarks undergoes a hard scattering event
and the final state $b \, \bar{b}$ pairs are highly asymmetric in $p_T$. In gluon splitting, the $b \, \bar{b}$ pairs
are dominantly produced with an opening angle due to splitting and neither of the quarks undergos a hard scattering interaction.
This typically leads to $b \, \bar{b}$ pairs with a small opening angle and small $p_T$, or both with large $p_T$ but not back to back. Using cuts,
the different processes can be separated out and a
dominant region of  $q \, \bar{q},  g \, g \rightarrow b \, \bar{b}$ production can be isolated. One strategy is to use muonic tags of the $b \, \bar{b}$ decays from
the semileptonic $\bar{B} \rightarrow \mu \, \bar{\nu}_{\mu} X$ decays and $\bar{B} \rightarrow J/\psi \, X \rightarrow \mu^+ \, \mu^-  \,X$ decays to allow triggering and to improve the identification and discrimination of $b$ jets from backgrounds.
A proof of principle is given by an initial CMS study that utilizes the opening angle between the $b$ quarks to distinguish
production mechanisms  \cite{CMS2}.

This initial study demonstrates the promise of discriminating the various production mechanisms in a hadron collider environment.
Note that \cite{CMS2} insisted on a single $\bar{B} \rightarrow J/\psi \, X \rightarrow \mu^+ \, \mu^-  \,X$ decay as
it was interested primarily in identifying states with small opening angles. We are interested in $b$ jets
that are approximately back to back with semi-leptonic muon decays. We advocate a discrimination of the production mechanisms
through identified $b$ jets
by enforcing cuts on the $p_T$ asymmetry and the pseudorapidities of the reconstructed $b \, \bar{b}$ pair of the following form,
\vspace{-5pt}
\begin{itemize}
\item [(i)]  To isolate the signal region with a small $p_T$ asymmetry enforce $p_T^b/p_T^{\bar{b}} <  (1+ N \, E)$ where $E$ is the total error on the reconstruction of the quark's partonic $p_T$.
The particular value of $N$ is not required for our  study as the LO production processes we consider trivially satisfy this condition. An optimal value of $N$ can be obtained in a NLO study.
The error for the $p_T$ asymmetry depends on the decay products and
the reconstruction algorithm used;  for the muonic decay tags that we advocate, a CMS study \cite{CMSnote} finds a reconstruction error of $13 \%$ on $p_T$
of the $B$ hadron and fragmentation function errors on the total cross section are on the order of $9 \, \%$.
This cut efficiently reduces the NLO flavor excitation production.
\vspace{-5pt}
\item [(ii)] The background due to gluon splitting can be reduced
through insisting that the pseudorapidities of the $b$ jets lie on different sides of the beam line in the polar angle. This can be done through imposing a cut with $\theta(-\eta_b \, \eta_{\bar{b}})$
on a NLO calculation.
The produced $b\,\bar{b}$ pairs will not lie exactly back to back at LO due to the residual boost resulting from the asymmetry of the partonic momenta via PDF effects.
\end{itemize}

These conditions are trivially satisfied by the two-body BSM or QCD production processes for the $p_T$ cuts that we consider
while efficiently suppressing the large NLO background from gluon splitting and flavor excitation. Further evidence in support of the efficiency of
these analytic cuts in isolating the signal region at LHC is supplied by past Monte Carlo studies of the kinematic properties of the
various heavy flavor creation processes at  the Tevatron \cite{Field:2002da}.

We compare the effect of BSM bosons on $b \, \bar{b}$ production
in the regions with the highest $p_T$ that should be accessible in the CMS detector.
We restrict the $p_T$ of the final state $b \, \bar{b}$ quarks to generally lie in the range $1 \,  {\rm \, TeV} < p_T < 1.5 {\rm \, TeV}$.
The pseudorapidity range is the full geometric acceptance
of the CMS detector for $|\eta| < 2.4$ \cite{CMSnote}.
It turns out that for such values of $p_T$ and $\eta$ those two quantities are independent from each other and we can impose separate cuts on them.\footnote{One can derive in a straightforward manner from Eq. (2) that in order to be able to impose independent cuts on pseudorapidity $\eta$ and $p_T$ it is sufficient to take $p_T > \sqrt{s}/ e^{\eta} $. For $|\eta| < 2.4$ this yields $p_T \gtrsim \sqrt{s}/11$.}
The differential cross section of the QCD and BSM processes in terms of the spin and color averaged matrix element
$\mathcal{M}$ (in terms of partonic Mandelstam variables) is given by
\bea\label{eq}
\frac{d\sigma}{d\eta \, d p_T^2} =  \int_0^1 d x_2  \int_0^1 d x_1 \frac{\langle|\mathcal{M}|^2\rangle \, \delta[x_1 \, x_2 \, \sqrt{s} - p_T \, ( x_1 e^{- \eta} + x_2 \,e^{\eta})]}{16 \, \pi x_1 \, x_2 \, s \, \sqrt{s}} \, f(x_1, \, \mu)  \, f(x_2, \, \mu)\ .
\eea
The $ f(x_i, \, \mu) $ are the PDFs evaluated at
the renormalization scale $\mu = m_b$. As usual, $x_i$ relate the partonic and hadronic momenta so that $P_{1,2} = x_{1,2} \, P^h_{1,2}$.
Integrating over pseudorapidity, the relevant differential cross section for our $p_T$ cuts is given by
\bea
\frac{d\sigma}{d p_T^2} =  \int_{\frac{4 p_T^2}{s}}^1 d x_2  \int_{\frac{4 p_T^2}{s \, x_2}}^1 d x_1 \frac{\langle|\mathcal{M}|^2\rangle}{8 \, \pi (s \, x_1 \, x_2)^2 \sqrt{1 - \frac{4 p_T^2}{s \, x_2  \, x_1}}}  \, f(x_1, \, \mu)  \, f(x_2, \, \mu).
\eea
 The LO QCD $2 \rightarrow 2$ amplitudes are given in Appendix A, section 1.
The results for various integrated luminosities, $p_T$ ranges, and CM energies for the LHC are given in
Table I.

\begin{table}
\begin{tabular}{|c||c|c|c|}
\hline
$\sqrt{s} \, \,  [{\rm TeV}]$ &  $\ \ \ \ \ {\int {\cal L} \, dt}\ \ \ \ \  $ & LO [$ \sqrt{s}/11$, 1.4]  & LO [1.0, 1.4]   \\
\hline \hline
 & 200 ${\rm pb^{-1}}$ & 68 & 4.4  \\ \cline{2-4}
\raisebox{2.5ex}[0pt]{7}  & 10 ${\rm fb^{-1}}$& $3.4 \times 10^3$  & 220   \\ \cline{2-4}
\hline
10  & 10 ${\rm fb^{-1}}$ & $2.3 \times 10^3$ & $1.3 \times 10^3$  \\ \cline{2-4}
\hline
14  & 10 ${\rm fb^{-1}}$ & $0.7 \times 10^3$ & $-$  \\ \cline{2-4}
\hline
\end{tabular}
\caption{Expected number of LO QCD partonic $b \, \bar{b}$ events for various integrated luminosities, $p_T$ ranges (in TeV), and operating energies at the LHC. We evaluate $\alpha_s$ at the
 $m_b$ scale throughout this paper and we use the central MSTW 2008 PDFs \cite{Martin:2009iq}.
The event rates of the QCD production and the BSM scenarios we consider will be both further suppressed
by common branching fractions to final states containing at least one muon. One should also apply a total event selection efficiency when the decay of a $b$ quark produces a muon of about $\sim 6.25 \%$, following \cite{CMSnote}.
These two factors will reduce the BSM signal and QCD background rate by a common factor of $\sim 10^{-4}$ and $\mathcal{O}(100) \, \rm {fb^{-1}}$ of data will be required when the LHC is operating at $\sqrt{s} \gtrsim 10 \, {\rm TeV}$ for a significant
QCD signal rate. As we show, less data can be required for a BSM signature; LHC operating at $\sqrt{s} \gtrsim 7 \, {\rm TeV}$ with integrated luminosity of $1 \, \rm fb^{-1}$  can find a clear signature of
BSM physics in the tail of the $p_T$ distribution in some models.}
\label{table1}
\end{table}

The contributions of the two-body QCD production mechanism for $\sigma(p \, p \rightarrow b \, \bar{b})$ up to NLO are known \cite{Gluck:1977zm,Owens:1977sj,Combridge:1978kx,Combridge:1977dm,Georgi:1978kx,Nason:1989zy,Ellis:1985er,Nason:1987xz}.
In comparing our BSM results with the SM QCD background we restrict ourselves to LO QCD for this production process.
We also restrict our attention to LO QCD due to collinear singularities in the NLO QCD production processes that are regulated by $m_b$.
We present results of one-loop interference calculations between the SM and BSM physics
where the $b$ quark mass is neglected.
To compare these results consistently with NLO QCD calculations we would need to include final state masses in both calculations as well as contributions from processes
with arbitrarily soft extra gluon emission.  The dominant NLO QCD production mechanisms that are larger than the LO QCD production mechanisms (flavour excitation and gluon splitting)
that we are concerned with are suppressed by the cuts we advocate. However, NLO QCD perturbative corrections of the form of virtual corrections
and initial and final state radiation corrections to the LO QCD production
rate, that do not have these distinct kinematic signatures, will not be highly suppressed by these cuts. Including such corrections
is beyond the scope and purpose of this work.\vspace{-7pt}\footnote{See \cite{Czakon:2008ii} for the status of analytic NLO QCD heavy flavor production and \cite{Beneke:2009ye} for a discussion of the status of the NNLO program.}

\vspace{10pt}
\section{Tree level exchanges of new bosons}
We treat the spectrum of the BSM physics
as an EFT with a single $\rm SU(2)_2$ singlet or doublet of bosons added to the SM in each case, assuming the cut-off scale is
sufficiently high compared to this mass so that higher dimensional operator effects can be neglected. We begin with the case with tree level
exchanges of a massive new spin one $\rm SU(3)_c$ boson with a purely vector coupling, sometimes referred to as a coloron in the literature.
\vspace{-16pt}
 \subsection{Colorons}
\vspace{-4pt}
Consider supplementing the SM with a massive spin one $\rm SU(3)_c$ octet
within the spectrum of BSM physics. Many models of dynamical electroweak symmetry breaking
with new strong interactions have such a particle, for details see \cite{Preskill:1980mz,Hall:1985wz,Frampton:1987dn,Hill:1991at,Hill:1993hs}.
A sufficient condition for this particle to appear in a BSM spectrum is to have a symmetry breaking pattern where a Higgs mechanism is employed
for QCD to emerge as a diagonal subgroup $\rm SU(3)_1 \times SU(3)_2 \rightarrow SU(3)_c$.
When the resultant spectrum is reduced to its mass eigenstate basis one obtains $\rm QCD$ with the addition of
a massive $\rm SU(3)_c$ octet. Gauge invariance forbids the single coupling of such a particle to gluon pairs
and pair production leading to four $b$ and four jet signatures of these states has been recently examined in \cite{Dobrescu:2007yp}.
Consider the case where the coupling of the massive $\rm SU(3)_c$ octet has purely vector-like flavor universal coupling
and $\mathcal{L}  \supset  \, g_3 \, \cot{\theta} \, G_{a}^\mu \, \bar{q} \, T^a \gamma_\mu \, q$, where $\cot{\theta} = \xi_2/\xi_1$ is the ratio of the
gauge couplings of the underlying $\rm SU(3)_1 \times SU(3)_2$ gauge groups.
Then single production of these states from quarks in $p\, p$ collisions will give rise to the
dominant source of $b \, \bar{b}$ pairs with the kinematics we are interested in. The spin and color averaged matrix element for the excess contribution
compared to QCD in the $s$-channel in terms of partonic Mandelstam variables is
\bea
\langle|\mathcal{M}|^2\rangle  = \frac{g^4}{4} \, (\cot{\theta})^4 \, C_F^2 \, \left(\frac{1}{(\hat{s} - m_c^2)^2 + m_c^2 \, \Gamma^2}+\frac{2 (\hat{s} - m_c^2) }{[(\hat{s} - m_c^2)^2 + m_c^2 \, \Gamma^2] \, \hat{s}} \right) \left[\hat{t}^{\,2} + \hat{u}^2 \right],
\eea
where the width of the coloron is approximated by its quark decays as $\Gamma \simeq \alpha_s   (\cot{\theta})^2  m_c$.
For colorons of mass $m_c \gtrsim 2 \, {\rm TeV}$ the excess of events in the $p_T$ tail is far larger than the uncertainties of the LO QCD estimate of the background.

We define a $5 \, \sigma$ significance by treating the $\rm LO$ QCD rate as a Poisson distribution. To estimate the
effect of large $\rm NLO$ corrections we multiply the $\rm LO$ QCD rate used in the Poisson distribution by $\left[1 + \frac{\alpha_s(m_b)}{4\pi} \, \ln \left(p_T^2/m_b^2\right)\right]$ and we take
$p_T = 1.5 \, {\rm TeV}$.\footnote{One can reduce the QCD errors through a $\rm NLO$ QCD calculation utilizing the $\rm FONLL$ formalism of \cite{Cacciari:1993mq}
which resums the large logs of $p_T/m_b$ but this is beyond the scope of this work. Such a resummation while running down from the $p_T$ scale to $m_b$ justifies our
choice of the renormalization scale at $\mu \simeq m_b$.} We insist on five signal events after a suppression factor of $10^{-4} \simeq (3 \%)^2 \times (25 \%)^2$ is applied for the branching ratios of both $b$ quarks
to final states containing a muon and a total signal efficiency of $25 \, \%$ \cite{CMSnote,CMS2} for each muon, $b$ jet trigger.

A $5 \, \sigma$ discovery of colorons is possible in this channel with a reach out to $\,\sim 2 \, p_T^{\rm max}$, however, a significant event rate will require $\gtrsim 1 \, \rm fb^{-1}$ of integrated luminosity because of the suppression of the signal due to selection efficiencies and branching ratios.
We show the total integrated luminosity required for various coloron masses and operating energies that pass these tests in Figure 2.
Other sources of events from the $t \, \bar{t}$ background in the signal region
are estimated to be a small $\%$ level background \cite{CMSnote}. Fakes from $c$ quarks are reduced with the associated $b$ jet trigger and light quark fakes are removed
by a $p_T > 15 \,{\rm GeV}$ cut on the muons. The experimental cross section uncertainty is approximately $20 \, \%$  \cite{CMSnote} for the $p_T$ regions we consider
and is dominated by a systematic jet energy scale uncertainty of $12 \%$.
\vspace{0.5cm}
\begin{figure}[h] \label{LOQCD}
\begin{center}
{\includegraphics[height=5.4cm]{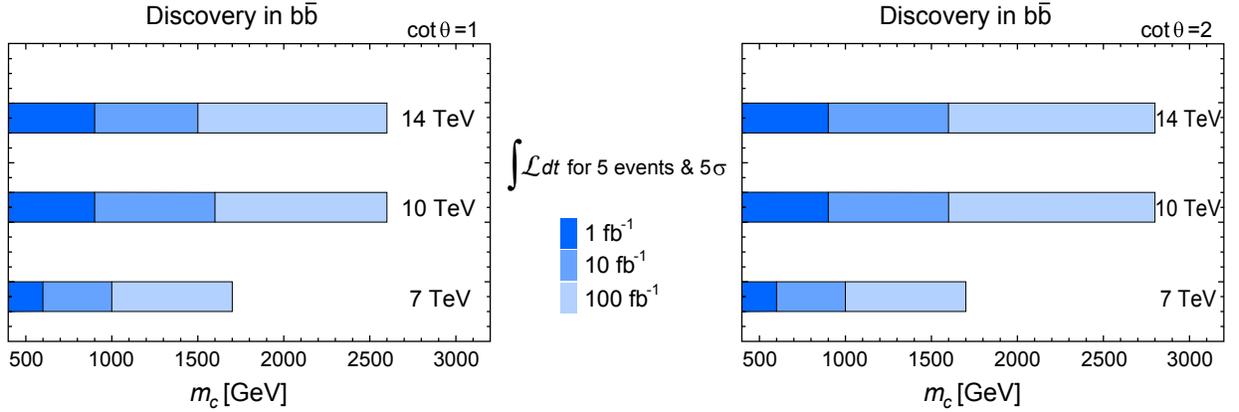}}
\caption{The discovery potential of colorons in the $p_T$ tail of $b \, \bar{b}$ production for $p_T$ in the range $1.0 - 1.4 \, \rm TeV$ for CM energies $7 \,\rm TeV$, $10 \,\rm TeV$, and in the range $\sqrt{s}/11$ to $1.4 \,\rm TeV$ for $ 14 \,\rm TeV$. We
show the range of coloron masses that can be discovered with $5 \, \sigma$ significance and at least 5 signal events after branching ratio and event selection suppression of $10^{-4}$ is applied to the signal rate. On the left we take $\cot{\theta} = 1$, on the right $\cot{\theta} = 2$. Our discovery reach is dominated by the requirement of $5$ signal events after the suppression of $10^{-4}$ is applied to the signal rate, not PDF or theory uncertainties. For example, the signal to background ratio for the entire discovery reach region for $\cot \theta = 1$ and $\sqrt{s} = 7 \, {\rm TeV}$ is $S/B > 98$.}
\end{center}
\end{figure}

Some of the coloron masses which the tail of the $p_T$ distribution is sensitive to are not ruled out by Tevatron studies.
The most stringent and most model-independent bounds on the mass scale of flavor universal
colorons come from resonance searches in the di-jet invariant mass differential distribution \cite{Aaltonen:2008dn} which obtains a $95 \%$ confidence limit exclusion bound on the
coloron masses in the range of $260 < m_c < 1250 \,{\rm GeV}$.  We emphasize the complementary nature of searches in this channel to the traditional search strategies for colorons that have focused on $t \,\bar{t}$ pair production.

\section{One-loop interference effects of BSM bosons and QCD}

New $\rm SU(3)_c$ bosons with tree level couplings to quarks, where $g \sim g_s$, can lead to a
large excess of events in the tail of the $d \sigma(p \, p \rightarrow b \, \bar{b})/d p_T^2$ distribution. For small widths a
resonance peak can also be reconstructed. However, when the tree level coupling of a new boson to quarks is proportional to the quark mass, which is frequently the case for
spin zero bosons, the largest contribution to high $p_T$ events can come from one-loop production interference effects with the SM LO production processes.
This is the case for a $(8,2)_{1/2}$ scalar representation that couples to quarks proportionally to the Yukawa coupling.  In this section we explore the possible effects on the
tail of the $d \sigma(p \, p \rightarrow b \, \bar{b})/d p_T^2$ distribution due to such a scalar in one-loop corrections.

One-loop effects should not be assumed too small to be detected for heavy flavor production $ab \, \, initio$ due to our kinematic isolation
of a signal region with a suppressed QCD background in the tail of the steeply falling $d \sigma(p \, p \rightarrow b \, \bar{b})/d p_T^2$ distribution.
A direct example of the need of considering naively higher order corrections that experience kinematic enhancements is given by the dominance of NLO QCD corrections over LO QCD in heavy flavor production
when all phase space is integrated over.

In this particular model, two classes of corrections are of potential interest:  a class of one-loop corrections to $g \, g \rightarrow b \, \bar{b}$ production that are proportional to an enhanced top Yukawa coupling and a separate class experiencing
a resonant enhancement in the $s$-channel. We investigate these loop effects in detail in the following subsection.
\vspace{-10pt}
\subsection{Octet  $(8,2)_{1/2}$  scalars}

In the Manohar-Wise model \cite{Manohar} one adds a single  $(8,2)_{1/2}$  scalar representation to the SM.
This representation is the only flavor singlet  \cite{Arnold:2009ay} scalar representation allowed by MFV
other than the Higgs. We denote the doublet as
\vspace{-10pt}
\bea
{S}^A = \left(\!\begin{array}{c} {S^+}^A \\
{S^0}^A \end{array} \!\right) \ ,
\eea
where $A$ is the color index.
The Yukawa sector is determined up to overall complex normalization constants, $\eta_U$ and $\eta_D$, as required by MFV to be
\bea
L = -\eta_U g_{i j}^U \bar{u}_{R \, i} T^A Q_j S^A-\eta_D g_{i j}^D \bar{d}_{R \, i} T^A Q_j S^A + h.c.\ ,
\eea
where $g^U$ and $g^D$ are the standard model Yukawa matrices and $i,j$ are flavor indices. See \cite{Burgess:2009wm} for a discussion on
the current status of the model's phenomenology.  The largest effects of this model on the tail of the $d \sigma(p \, p \rightarrow b \, \bar{b})/d p_T^2$ distribution do not come from tree level exchanges.
This is due in part to the suppression by light quark Yukawa couplings. The $b,t$ quark Yukawa couplings are far larger but their anti-quark PDFs are highly suppressed in the proton. Further, gauge invariance
forbids a coupling of a single octet scalar to two gluons. The largest effects on $d \sigma(p \, p \rightarrow b \, \bar{b})/d p_T^2$ come from two
potential sources: one-loop production mechanisms that afford an $s$-channel resonance with an ${S^0}$ exchange, and one-loop interference terms with
the QCD production mechanisms which are proportional to $\left(\!\frac{m_t}{v}\!\right)^2$ due to $S^{\pm}$ loops.
We discuss the effect of each of these on the tail of the $d \sigma(p \, p \rightarrow b \, \bar{b})/d p_T^2$ distribution in turn.

The largest $s$-channel resonant production is shown in Figure 3.
\begin{figure}[h]
\begin{center}
{\includegraphics[height=5.5cm]{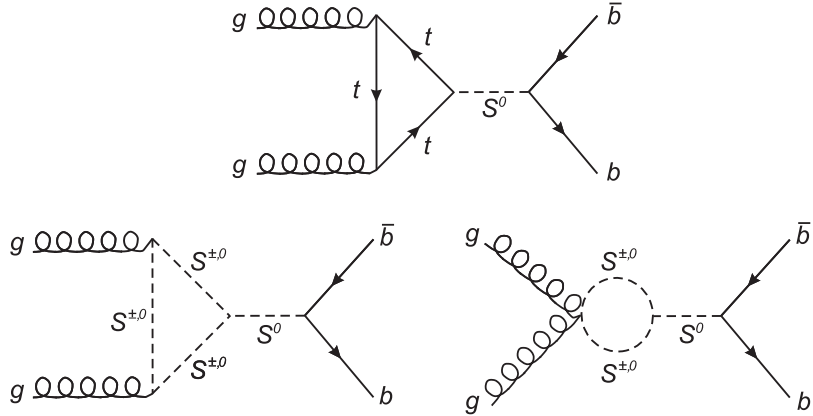}}
\caption{One-loop production of neutral  $(8,2)_{1/2}$  scalars.}
\end{center}
\end{figure}
\vspace{-15pt}
From the loop parts of those diagrams, the biggest effect on single production of neutral scalars  comes from the top quark loop \cite{Gresham:2007ri}.
This is also true for the $s$-channel exchange and we neglect the effects of the other diagrams that are also proportional to the
unknown couplings in the scalar potential of the model ($\lambda_{4,5}$), which are expected to further suppress the contribution of the scalar loop diagrams. The amplitude for the top loop is given by
\bea
i \mathcal{M}_s =    \frac{i \,m_t^2 \, m_b}{8 \, \pi^2v^2} \, \frac{\eta_U \, \eta_D \, g_s^2\, d^{ABC}\, \bar{u}_b \, T^C \, v_b}{s - m_s^2 + i \, m_s \, \Gamma} \int_0^1 dx \int_0^{1-x} dy \left[\frac{(p_1 \cdot \epsilon_2 \, p_2  \cdot \epsilon_1 - \epsilon_1 \cdot \epsilon_2 \,  p_1 \cdot p_2)(1 - 4 x y)}{m_t^2 - s x y} \right]
\eea
where we have taken $\eta$'s real and neglected terms proportional to $p^1_\mu, p^2_\nu$ as discussed in the Appendix.
We approximate the width of the real scalar  by $\Gamma \simeq \frac{m_s}{16\pi}\left(\frac{m_t}{v}\right)^2 \eta_U^2\left(1-\frac{4m_t^2}{m_s^2}\right)^{3/2}$  as it is dictated by top decays \cite{Manohar}.
We have determined $|\mathcal{M}_{SM}^{gg}+\mathcal{M}_s|^2$ and used equation (3) to calculate the resulting contribution of the excess events in the $p_T$ tail. Unlike the SM Higgs, as the $S^0$ is an 8 under $\rm SU(3)_c$, the amplitude in equation (7) interferes with the
SM heavy flavor production amplitudes given in Appendix A, section 1. This process leads to the change in the number of events
given by $\Delta E_{res}^{(2)}$ in Table II and the interference introduces another factor of the quark mass.
The resulting event rate is too small to be observed
because the resonant enhancement is suppressed by $\frac{m_b}{v}$ and loop factors.\footnote{
We do expect the corresponding resonance signal for $t$ quark final states to have an observable event rate
in the $p_T$ tail of the $d \sigma(p \, p \rightarrow t \, \bar{t})/ d p_T^2$ distribution.}
For large masses the parameters $\eta_U, \eta_D$ can be greater than one, however, the constraints from this model's loop corrections to
$b \rightarrow s \, \gamma$ and $R_b$ (see, \cite{Manohar,Gresham:2007ri}) still constrain $\eta_U, \eta_D < 10$, so
no statistically significant excess is observable in the $p_T$ tail from these processes with final state $b$ quarks.

The largest non-resonant effect comes from the interference of one-loop corrections due to $S^\pm$ with the SM LO production processes. These corrections are proportional to $\left(\!\frac{m_t}{v}\!\right)^2$.
The resulting amplitudes and diagrams are given in the Appendix. Again, after a rather lengthy calculation we have determined the excess event rate due to these effects using equation (3).
We summarize in Table II the numbers of expected excess events for the various loop processes.

\begin{table}
\begin{tabular}{|c|||c|c|c|c|c|}
\hline
$m_s \,  [{\rm GeV}]$ & $\ $\ 200\ \ $\ $& 500 & 1000  &  2000 & 3000  \\
\hline \hline
$\Delta  E(q \,\bar{q} \rightarrow b \, \bar{b})/|\eta_U|^2$ & 1.3 & -0.2 & -1.1 & -0.2 & -0.05\\
\hline
$\Delta  E(g \, g \rightarrow b \, \bar{b})/|\eta_U|^2$ & 51 & 38 & 18 & 3.3 & 1.7 \\
\hline
$\Delta  E_{res}^{(1)}(g \, g \rightarrow b \, \bar{b})/|\eta_U \eta_D|^2 \, $ & $-$ & $-$ & $\sim10^{-4}$ &$\sim10^{-3}$  & $\sim10^{-4}$ \\
\hline
$\Delta  E_{res}^{(2)}(g \, g \rightarrow b \, \bar{b})/|\eta_U \, \eta_D| \, $ & $-$ & $\sim10^{-3}$ & $\sim10^{-4}$ & $\sim$-$10^{-3}$& $\sim10^{-4}$ \\
\hline
$R_b^{1 \sigma} \, \rightarrow |\eta_U| <$ & 0.3 & 0.5 & 0.8 & 1.3 & 1.8  \\
\hline
$\delta(b \rightarrow s \, \gamma) \, \rightarrow |\eta_U \, \eta_D|<$ & 0.1& 0.4 & 0.9& 2.4 & 4.5\\
\hline
\end{tabular}
\caption{The change in the number of $b \, \bar{b}$ events due to interference with the octet amplitudes where
we restrict  $p_T$ between 1.0 and 1.4 $\rm TeV$, and use an operating energy of $\sqrt{s} = 10 \, {\rm TeV}$ for LHC.
The results in the table do not have a selection cut or branching ratios to muons applied.
The results are given using MSTW PDFs for $10 \, \rm fb^{-1}$ of integrated luminosity. We use  $m_t = 173.1 \, {\rm GeV}$, $v = 246 \, \rm GeV$
and take $\alpha_s(m_b) = 0.22$.
The process for $\Delta  E(q \,\bar{q} \rightarrow b \, \bar{b})$ is given in  Appendix A, section 2.
The process for $\Delta  E(g \, g \rightarrow b \, \bar{b})$ is given in  Appendix A, section 3. We have checked that the processes for
$\Delta  E(q \,\bar{q} \rightarrow b \, \bar{b})$ and $\Delta E(g \, g \rightarrow b \, \bar{b})$ analytically have the correct decoupling behavior.
The remaining two processes come from the direct square and the interference of equation (7)
with the LO QCD results in Appendix A, section 1. These are negligible despite the resonance enhancement. We present the results with the unknown $\eta_U$ and $\eta_D$ factored out. We also show the constraints
on these parameters from allowing a  $1 \sigma$ deviation of $R_b$ \cite{Gresham:2007ri} and a $10 \%$ deviation in $b \rightarrow s \, \gamma$  \cite{Manohar}. Light octets are not ruled out by
direct production and are consistent with EWPD down to $\sim 100 \, \rm GeV$ \cite{Burgess:2009wm}. All of these mechanisms do not lead to a statistically significant
excess of events in the $p_T$ tail of $b\,\bar{b}$  production.}
\label{table2}
\end{table}

For the $p_T$ range of integration $1.0 - 1.4 \ \rm TeV$ the results are not large enough to be detected,
despite the highly suppressed background and the enhancement of these loop effects.
The LO QCD background to compare these results to is given in Table I. The experimental uncertainty of the cross section in the high $p_T$ region has a systematic uncertainty on the
order of $20 \%$. No statistically significant excess is expected in this model in the $p_T$ tail. Our results additionally imply that these effects on the $b \, \bar{b}$ cross section determined at the Tevatron are also negligible,
affording no significant constraint on the model.

\vspace{1cm}

\section{Conclusions}
Heavy flavor production at hadron colliders has a long history of surprises and challenges to both theorists and experimentalists.
Due to the long standing difficulties of accommodating the measured $\sigma(p \, p \rightarrow b \, \bar{b})(p_T>p_T^{\rm cut})$ at the Tevatron
within our understanding of QCD, which have only recently been overcome, exploring  high $p_T$ production of $b\,\bar{b}$ pairs at the LHC
as a constraint on BSM physics demands great caution.  Despite this, one can impose simple kinematic cuts to
improve the discovery reach of heavy flavor production by suppressing the NLO QCD flavor excitation and gluon splitting processes.
This allows one to more efficiently exploit $b$ tagging with associated muons which affords a significant $p_T$ reach
for final state $B$ mesons at LHC.

We have determined the reach for a $5 \,\sigma$ discovery of flavor universal colorons using this technique at LO in QCD.
We have also demonstrated that the effects on $b \, \bar{b}$ production in the Manohar-Wise model are not large enough to afford a statistically
significant deviation from the expected QCD production background.

For the $p_T$ reach in $d \sigma(p \, p \rightarrow b \, \bar{b})/d p_T^2$ to be fully exploited in the LHC era, further work on precision studies of heavy flavor production
in the standard model is of increased importance in light of this discovery potential.  Improvements of the SM calculations of heavy flavor production
and the associated systematic uncertainties would have to be carefully examined at LHC before any evidence of
new physics could be claimed. However, our results are promising for further work
to develop the heavy flavor production program at LHC with this aim for discovery of new massive $\rm SU(3)_c$ bosons with tree level couplings to quarks
not proportional to the quark mass.

\vspace{1cm}

\subsection*{Acknowledgment}
We thank Mark Wise for collaboration during the early stages of the work presented here.
We further thank Mark Wise, Maxim Pospelov, Saba Zuberi and Brian Batell for comments on the manuscript.
Research at the Perimeter Institute is supported in part by the Government of Canada through NSERC and by the Province
of Ontario through MEDT.

\vspace{1cm}

\appendix
\section{QCD and $(8,2)_{1/2}$ octet results}
\subsection{LO QCD results}

The LO and NLO QCD results for $\sigma(p \, p \rightarrow b \, \bar{b})$ are well known in the literature \cite{Gluck:1977zm,Owens:1977sj,Combridge:1978kx,Combridge:1977dm,Georgi:1978kx,Nason:1989zy,Ellis:1985er,Nason:1987xz}.
We summarize the simple LO results for completeness and to explain a trick \cite{Feynman} that we employ to simplify our
subsequent one-loop calculations. As discussed in Section II,
cuts on the $p_T$ asymmetry and the opening angle of the reconstructed $b \, \bar{b}$ pair
can isolate the kinematic region where the dominant QCD (or BSM) production mechanisms
are the  $q \, \bar{q},  g \, g \rightarrow b \, \bar{b}$ processes. The LO results for these processes in QCD are given by
\begin{eqnarray}
i  \mathcal{M}^{q\bar{q}}_{SM}&=&-i g_s^2 \, \frac{1}{s} \left[\bar{v}(p_2) \, \gamma^{\mu} \,  T^A \, u(p_1)\right] \, \left[\bar{u}(p) \, \gamma_{\mu} \, T^A \, v(\bar{p}) \right], \nonumber\\
i  \mathcal{M}^{gg}_{SM}&=&-i g_s^2\, \bar{u}(p)\bigg[[T^A,T^B]\frac{1}{s} \left(g^{\mu\nu}(\slashed{p}_1-\slashed{p}_2)+2\gamma^{\nu}p_2^{\mu}-2\gamma^{\mu}p_1^{\nu}\right)\nonumber\\
&&\ \ \ \ \ \ \ \ \ \ \ \ \ \ \ \ +\,T^BT^A\frac{1}{t} \gamma^{\nu}(\slashed{p}_1-\slashed{\bar{p}})\gamma^{\mu}+
T^AT^B\frac{1}{u}\gamma^{\mu}(\slashed{p}-\slashed{p}_1)\gamma^{\nu}\bigg]
v(\bar{p})\ .
\end{eqnarray}

Note that in $i  \mathcal{M}^{gg}_{SM}$ we have modified the triple gluon vertex of QCD to eliminate the
dependence on $p_1^\mu,p_2^\nu$. This does not change the final answer once the gluon is summed over physical polarizations
as the difference is proportional to $p_1^\mu \cdot \epsilon_\mu(p_1)$ and $p_2^\nu \cdot \epsilon_\nu(p_2)$ \cite{Georgi:1978kx}.
This trick does, however, allow us to sum over the physical and non-physical polarizations of the gluon as the spurious contributions
to the amplitude, which are proportional to this momentum dependence, are removed. When calculating the interference
of the LO QCD processes with such a modification, this trick allows one to set to zero all factors of $p_1^\mu,p_2^\nu$ in our results for
the one-loop interference effects of the $(8,2)_{1/2}$ scalar with QCD, and sum over all polarizations of the gluon, simplifying significantly intermediate steps of the calculation.

\subsection{Octet effects on $q\, \bar{q} \rightarrow b \, \bar{b}$}
Our one-loop result for the modification of $i  \mathcal{M}^{q\bar{q}}_{SM}$ (see Figure 4) is given as the following gauge invariant amplitude (where we have set all external masses to zero)
\begin{eqnarray}
i  \mathcal{M}_O^{q\bar{q}} &=& \frac{i\,|\eta_u|^2\, g_s^2}{16 \, \pi^2 \, s}   \left(\frac{m_t}{v}\right)^2 \left[\bar{u}_b(p) \, \gamma^\mu \, T^A \, P_L \, v_b(\bar{p})\right] \,  \left[\bar{v}_i(p_2) \, \gamma_\mu \, T^A u_i (p_1) \right]\nonumber\\
&&\times \Bigg[C_A \bigg[
A_1(m_s^2,m_t^2) -A_1(m_t^2,m_s^2) + A_2(m_s^2,m_t^2)+ \frac{1}{2} \bigg]\nonumber\\
&&+ \,\,C_F \bigg[\,
2 \, A_1(m_t^2,m_s^2) - \,2 \, A_2(m_s^2,m_t^2) + \frac{3m_t^2-m_s^2}{2(m_s^2-m_t^2)}\,\bigg]\nonumber\\ && + \,\,C_F\bigg[\,\frac{m_s^2(m_s^2-2m_t^2)}{(m_s^2-m_t^2)^2}\ln\left(\frac{m_s^2}{\mu^2}\right) +
\frac{m_t^4}{(m_s^2-m_t^2)^2}\ln\left(\frac{m_t^2}{\mu^2}\right)
\bigg]\Bigg] \ ,
\end{eqnarray}
where
\bea
A_1(m_s^2,m_t^2) &=& \int_0^1 \, dx \, \frac{m_{s}^2 x - m_t^2 (x-1)}{m_{s}^2 - m_t^2 - s x} \, \ln \left(\frac{m_s^2 x - m_t^2(x-1) }{\mu^2} \right)  \nn \\
&\,&-\int_0^1 \, dx \, \frac{m_{s}^2 + s (x-1) x}{m_{s}^2 - m_t^2 - s x} \, \ln \left(\frac{m_s^2 + s (x-1) x}{\mu^2} \right), \nn \\
A_2(m_s^2,m_t^2) &=&  - \int_0^1 \, dx \, \frac{s (x-1) \, x}{m_s^2 - m_t^2 + s x} + \frac{m_t^2}{m_s^2 - m_t^2 + s x} \, \ln \left(\frac{m_s^2 (1-x) + x \, m_t^2}{m_t^2 + s (x-1) x} \right) \nn \\
&\,&  +  \int_0^1 \, dx \,  \frac{s x(m_s^2 (x-1) - m_t^2 x)}{(m_s^2 - m_t^2 + s x)^2} \, \ln \left(\frac{m_s^2 (1-x) + x \, m_t^2}{m_t^2 + s (x-1) x} \right),
\eea
and for the $\rm SU(3)_c$ case $C_A=3, C_F=\frac{4}{3}$.
\vspace{1cm}
\begin{figure}[h] \label{octetqq}
\begin{center}
{\includegraphics[height=6.0cm]{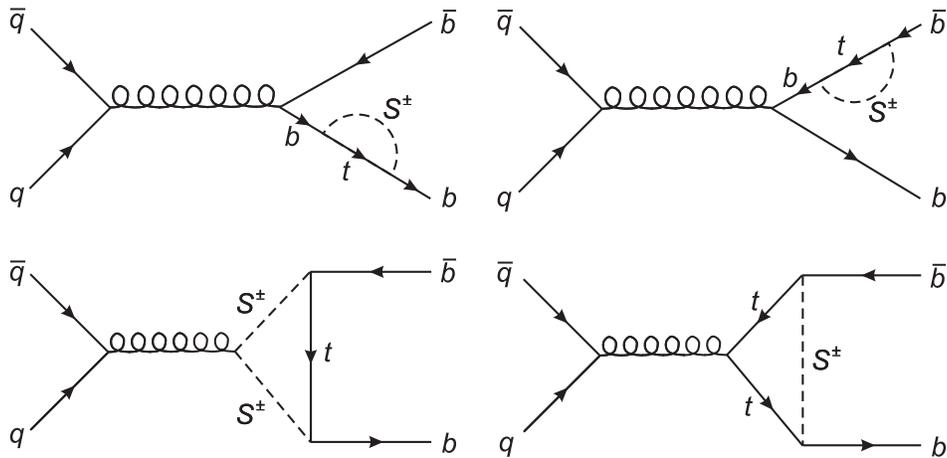}}
\caption{One-loop effects of the $(8,2)_{1/2}$ scalar on $q\, \bar{q} \rightarrow b \, \bar{b}$. The relevant cross section is directly obtained through interference
with the SM process. }
\end{center}
\end{figure}

\subsection{Octet effects on $g\, g \rightarrow b \, \bar{b}$}
Our one-loop results for the modification of $i \mathcal{M}^{gg}_{SM}$ (see Figure 5) are given as
\begin{eqnarray*}
i \mathcal{M}_{O}^{gg} &=& \frac{i\,g_s^2 \, |\eta_u|^2}{16 \, \pi^2} \, \left(\frac{m_t}{v}\right)^2 \left(\mathcal{M}^{t}_{a,b} + \mathcal{M}^t_{c} +  \mathcal{M}^t_{d}+ \mathcal{M}^t_{e} +
\mathcal{M}^t_{f}+ \mathcal{M}^t_{g}+ \mathcal{M}^t_{h}+ \mathcal{M}^t_{i}+ \mathcal{M}^t_{j}+ \mathcal{M}^t_{k}  \right. \nonumber\\
&&\left.+  \mathcal{M}^{u}_{a,b} + \mathcal{M}^u_{c} +  \mathcal{M}^u_{d}+ \mathcal{M}^u_{e} +
\mathcal{M}^u_{f}+ \mathcal{M}^u_{g}+ \mathcal{M}^u_{h}+ \mathcal{M}^u_{i}+ \mathcal{M}^u_{j}+ \mathcal{M}^u_{k} + \mathcal{M}^s_{l,m}+ \mathcal{M}^s_{n}+ \mathcal{M}^s_{o}\right).
\end{eqnarray*}
\begin{figure}[h] \label{octetgg}
\begin{center}
{\includegraphics[height=13.5cm]{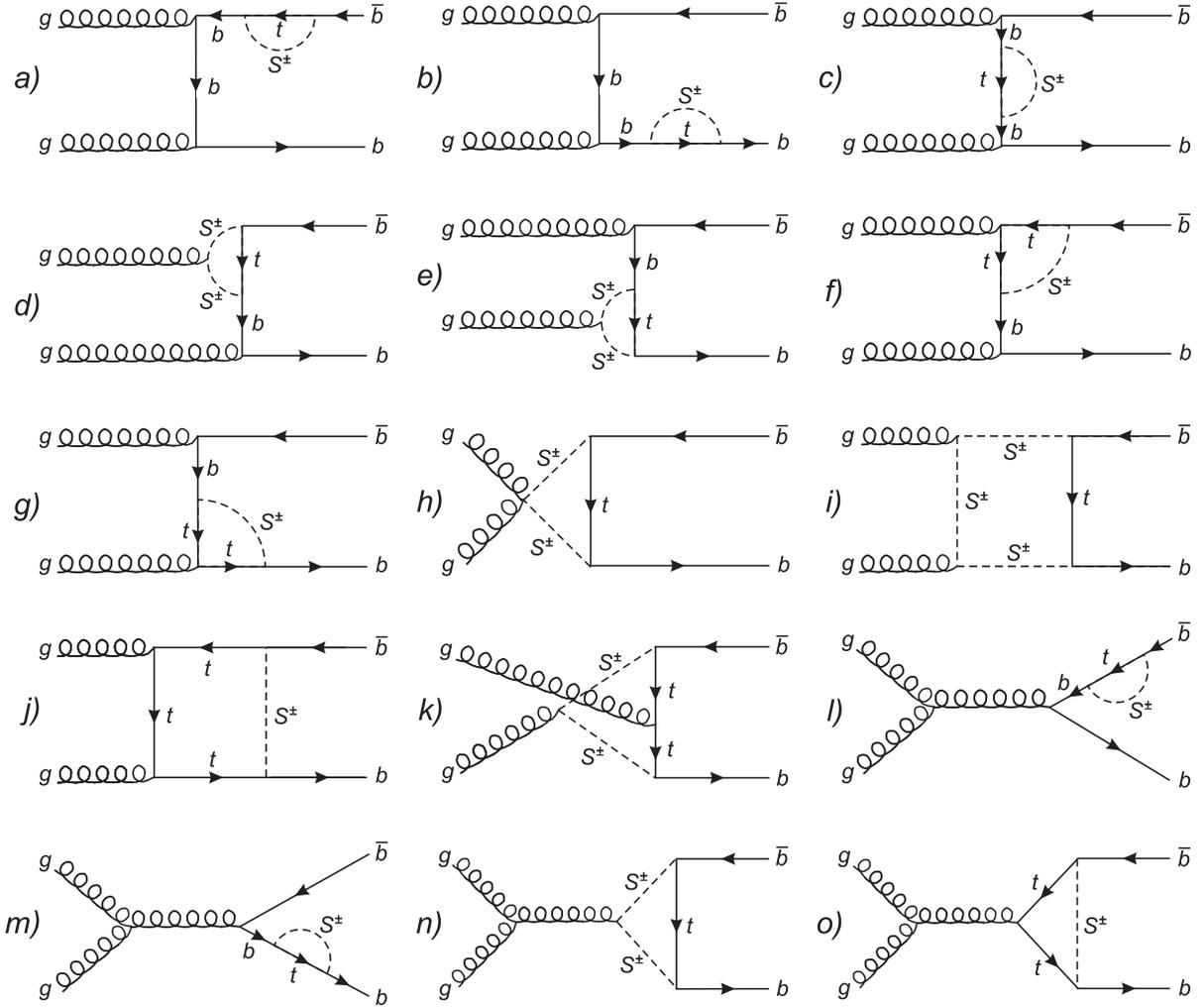}}
\caption{One-loop effects of the $(8,2)_{1/2}$ scalar on $g\, g\rightarrow b \, \bar{b}$. The relevant cross section is directly obtained through interference
with the SM process in a rather tedious and lengthy calculation. }
\end{center}
\end{figure}

\newpage

The formulas for the $u$-channel diagrams are obtained from the $t$-channel results by substituting $t\rightarrow u$.
We define the following Feynman parameter volumes as a short hand
\bea
\int d V(x,y) &=& \int_0^1 dx \int_0^{1-x} dy, \nn \\
\int d V(x,y,z) &=& \int_0^1 dx \int_0^{1-x} dy \int_0^{1-x-y} dz.
\eea
Our $t$-channel results are as follows (again, after setting all external masses to zero).
The expressions below should be contracted with $\epsilon_A^{\mu}(p_1) \, \epsilon_B^{\nu}(p_2)$.
\begin{eqnarray}
\mathcal{M}^t_{a,b}&=& -\frac{2 \, C_F \, T^B \, T^A}{t} \, \bar{u}_L(p) \gamma_{\nu}(\slashed{p}_1-\slashed{\bar{p}})\gamma_{\mu}v_L(\bar{p})\int_0^1 dx x\ln\left(\frac{m_s^2x+m_t^2(1-x)}{\mu^2}\right)\ ,
\nonumber\\
\mathcal{M}^t_{c}&=&
-\frac{2 \, C_F \, T^B \, T^A}{t} \,
\bar{u}_L(p) \gamma_{\nu}(\slashed{p}_1-\slashed{\bar{p}})\gamma_{\mu}v_L(\bar{p})\int_0^1 dx x\ln\left(\frac{m_s^2x+m_t^2(1-x)+t x(x-1)}{\mu^2}\right)\ ,
\nonumber\\
\mathcal{M}^t_{d}&=&\frac{ C_A \, T^B \, T^A}{t} \,
\Bigg[\bar{u}_L(p) \gamma_{\nu}(\slashed{p}_1-\slashed{\bar{p}})\gamma_{\mu}v_L(\bar{p}) \, \int dV(x,y) \, \ln\left(\frac{ F_1(x,y)}{\mu^2}\right)\nonumber\\
&&+ \int d V(x,y) \, \frac{1}{F_1(x,y)}
\bar{u}_L(p)\gamma_{\nu}v_L(\bar{p})[p_{1\mu}(2x-1)+2\bar{p}_{\mu}(1-x-y)]x t\Bigg]\ ,
\nonumber\\
\mathcal{M}^t_{e}&=&
\frac{ C_A \, T^B \, T^A}{t} \,
\Bigg[\bar{u}_L(p) \gamma_{\nu}(\slashed{p}_1-\slashed{\bar{p}})\gamma_{\mu}v_L(\bar{p})  \, \int d V(x,y) \, \ln\left(\frac{ F_2(x,y)}{\mu^2}\right)\nonumber\\
&&+ \, \int d V(x,y) \, \frac{1}{F_2(x,y)}
\bar{u}_L(p)\gamma_{\mu}v_L(\bar{p})[(p_1-\bar{p})_{\nu}(2x-1)+p_{\nu}(2y-1)]x t\Bigg]\ ,
\nonumber\\
\mathcal{M}^t_{f}&=&
\frac{\left( 2 \, C_F - C_A\right)T^BT^A}{t}
\Bigg[\bar{u}_L(p) \gamma_{\nu}(\slashed{p}_1-\slashed{\bar{p}})\gamma_{\mu}v_L(\bar{p})\,  \int d V(x,y) \,  \left[ \ln \left(\frac{F_{3}(x,y)}{\mu^2}\right)+1\right]
\nonumber\\
&&-
\int d V(x,y) \,  \frac{1}{ F_{3}(x,y)}\bar{u}_L(p)
\gamma_{\nu}(\slashed{p}_1-\slashed{\bar{p}})
[\slashed{p}_1(1-x)-\slashed{\bar{p}}y+m_t]\gamma_{\mu}[-\slashed{p}_1x - \slashed{\bar{p}}y + m_t]v_L(\bar{p})
\Bigg]\ ,
\nonumber\\
\mathcal{M}^t_{g}&=&
\frac{\left(2 \, C_F - C_A\right)T^BT^A}{t}
\Bigg[ \bar{u}_L(p) \gamma_{\nu}(\slashed{p}_1-\slashed{\bar{p}})\gamma_{\mu}v_L(\bar{p})\int d V(x,y) \, \left[\ln \left(\frac{F_{4}(x,y)}{\mu^2}\right)+1\right]\nonumber\\
&&-
\int d V(x,y) \,  \frac{1}{ F_{4}(x,y)}\bar{u}_L(p)
[\slashed{p}_2x+\slashed{p}y+m_t]\gamma_{\nu}[\slashed{p}_2(x-1)+\slashed{p}y+m_t]
(\slashed{p}_1-\slashed{\bar{p}})\gamma_{\mu}v_L(\bar{p})
\Bigg]\ ,
\nonumber\\
\mathcal{M}^t_{h}&=&0\ ,
\end{eqnarray}
where
\begin{eqnarray}
F_{1}(x,y)&=&m_s^2 (x+y) + m_t^2(1-x-y) + t x (x+y-1)\ ,\nonumber\\
F_{2}(x,y) &=& m_s^2(x+y) + m_t^2(1-x-y) + t x (x+y-1)\ ,\nonumber\\
F_{3}(x,y)&=&m_s^2 y + m_t^2(1-y) - t x y\ , \nonumber\\
F_{4}(x,y)& =& m_s^2 y + m_t^2(1-y) - t x (2-2x-y)\ .
\end{eqnarray}

The $s$-channel and $t$-channel box diagram results are
\begin{eqnarray}
\mathcal{M}^t_{i}&=&
-\,4 \,C_A\,\left( T^BT^A +  \frac{1}{2}\delta^{BA}\right)
\int d V(x,y,z) \, \nonumber\\
&& \times \Bigg[\bar{u}_L(p)\gamma_{\nu}v_L(\bar{p})\bigg[\frac{\bar{p}_{\mu}(1-x-z)+p_{\mu} y}{2F_{5}(x,y,z)}\bigg]
- \bar{u}_L(p)\gamma_{\mu}v_L(\bar{p})\bigg[\frac{p_{\nu}(1-x-y)+\bar{p}_{\nu}z}{2F_{5}(x,y,z)}\bigg] \nn \\
&&
+\,\bar{u}_L(p)\slashed{p}_1v_L(\bar{p})\bigg[ \frac{x g_{\mu\nu}}{2F_{5}(x,y,z)}+\frac{x\left[\bar{p}_{\mu}(1-x-z)+p_{\mu} y\right]
\left[p_{\nu}(1-x-y)+\bar{p}_{\nu}z\right]}{[F_{5}(x,y,z)]^2}\bigg] \, \Bigg],  \nn \\
\mathcal{M}^t_{j}\!\!&=&\!\!\!
\left[(2 C_F - C_A)T^BT^A  + \frac{C_A}{2}\delta^{BA}\right]\!\int \!d V(x,y,z)\Bigg[
\frac{1}{F_{6}(x,y,z)}\bar{u}_L(p) \gamma_{\nu}\gamma_{\mu}[\slashed{p}_1(y+z-1)-\slashed{p}_2 x]  v_L(\bar{p}) \Bigg. \nn \\
&& \Bigg.+\frac{1}{F_{6}(x,y,z)}\bar{u}_L(p)\Big[\gamma_{\mu}[\slashed{p}_1 z - \slashed{p}_2 (x+y)+\slashed{p}y]\gamma_{\nu}+
[\slashed{p}_1 z +\slashed{p}_2(1-x-y)]\gamma_{\nu} \gamma_{\mu}\Big]v_L(\bar{p}) \Bigg. \nn \\
&&\Bigg.+
\frac{1}{[F_{6}(x,y,z)]^2} \bar{u}_L(p) m_t^2 \Big[ [\slashed{p}_2(1-x-y)+\slashed{p}_1 z]\gamma_{\nu}\gamma_{\mu}
+ \gamma_{\nu}[-\slashed{p}_2 (x+y) + \slashed{p}_1 z + \slashed{p} y]\gamma_{\mu}\Big]v_L(\bar{p}) \Bigg. \nn \\
&&\Bigg.+
\frac{1}{[F_{6}(x,y,z)]^2} \bar{u}_L(p) m_t^2\gamma_{\nu}\gamma_{\mu}[-\slashed{p}_1 (1-y-z)-\slashed{p}_2 x]v_L(\bar{p}) \Bigg. \nn \\
&& \Bigg.\!\!\!\!\!\!\!\!\!\!\!\!\!\!\!\!\!\!\!\!\!+
\frac{1}{[F_{6}(x,y,z)]^2} \bar{u}_L(p) [\slashed{p}_2(1-x-y)+\slashed{p}_1 z ]\gamma_{\nu}[-\slashed{p}_2(x+y)+\slashed{p}_1 z + \slashed{p}y]\gamma_{\mu}[-\slashed{p}_2 x - \slashed{p}_1(1-y-z)] v_L(\bar{p})
\Bigg],
\nonumber\\
\mathcal{M}^t_{k}&=&
- 8 \, C_A \, \delta^{AB} \int d V(x,y,z)  \Bigg[
\frac{1}{F_{7}(x,y,z)} \, \bar{u}_L(p) \, \gamma_{\mu}[p_2^{\nu}(2y+2z-1)-2\bar{p}^{\nu}y+2p^{\nu}x]  v_L(\bar{p}) \Bigg. \nn \\
&& + \Bigg. \,\frac{1}{F_{7}(x,y,z)} \,  \bar{u}_L(p)\Big[\gamma_{\nu}\gamma_{\mu}[-\slashed{p}_2(y+z-1)-\slashed{p}x+m_t]
+[-\slashed{p}_2(y+z)+\slashed{\bar{p}}y+m_t]\gamma_{\mu}\gamma_{\nu}\Big]v_L(\bar{p}) \Bigg. \nn \\
&&+\,\frac{1}{[F_{7}(x,y,z)]^2}\bar{u}_L(p)[-\slashed{p}_2(y+z)+\slashed{\bar{p}}y+m_t]\gamma_{\mu}
[-\slashed{p}_2(y+z-1)-\slashed{p}x+m_t][2\bar{p}^{\nu}y-2p^{\nu}x]v_L(\bar{p})
\bigg],
\nonumber\\
\mathcal{M}^s_{l,m}\!&=&\!\!-\frac{2 C_F}{s} \, \left[T^A,T^B\right]
\bar{u}_L(p)\left[g^{\mu\nu}
(\slashed{p}_1-\slashed{p}_2)+2\gamma^{\nu}p_2^{\mu}-2\gamma^{\mu}p_1^{\nu}\right]v_L(\bar{p})
 \int_0^1 dx \, x \,  \ln \left(\frac{m_s^2x+m_t^2(1-x)}{\mu^2}\right),\nonumber\\
\mathcal{M}^s_{n}&=&\frac{C_A}{s} \left[T^A,T^B\right]
\bar{u}_L(p)\left[g^{\mu\nu}
(\slashed{p}_1-\slashed{p}_2)+2\gamma^{\nu}p_2^{\mu}-2\gamma^{\mu}p_1^{\nu}\right]v_L(\bar{p}) \int d V(x,y) \, \ln \left(\frac{F_8(x,y)}{\mu^2}\right),
\nonumber\\
\mathcal{M}^s_{o}&=& \frac{\left( 2 C_F - C_A\right)}{s} \, \left[T^A,T^B\right] \,
\bar{u}_L(p)\left[g^{\mu\nu}
(\slashed{p}_1-\slashed{p}_2)+2\gamma^{\nu}p_2^{\mu}-2\gamma^{\mu}p_1^{\nu}\right]v_L(\bar{p}) \nn \\
&&\times \int d V(x,y) \, \bigg[\ln \left(\frac{F_9(x,y)}{\mu^2}\right)+1- \frac{m_t^2+ s x y }{F_9(x,y)}\bigg]\ ,\!\!\!\!\!\!\!\!\!\!\!\!\!
\end{eqnarray}
where
\begin{eqnarray}
F_{5}(x,y,z)&=&m_s^2(x+y+z)+m_t^2(1-x-y-z)- s y z - t x (1-x-y-z)\ ,\nonumber\\
F_{6}(x,y,z)&=&m_s^2y+m_t^2(1-y)- s x z - t y (1-x-y-z)\ ,\nonumber\\
F_{7}(x,y,z)&=&m_s^2(x+z)+m_t^2(1-x-z)+s y z+t(x+y)(x+z)\ ,\nonumber\\
F_8(x,y)&=&m_s^2(x+y)+m_t^2(1-x-y)-s x y\ ,\nonumber\\
F_9(x,y)&=&m_t^2(x+y)+m_s^2(1-x-y)-s x y \ .
\end{eqnarray}



\end{document}